\documentclass[%
 reprint,
 amsmath,amssymb,
nofootinbib
]{revtex4-2}

\usepackage{graphicx}
\usepackage{dcolumn}
\usepackage{bm}
\usepackage{tikz}
\usetikzlibrary{quantikz}

\usepackage{quantikz}
\usepackage[utf8]{inputenc}
\usepackage{physics}
\usepackage{amsmath}

\usepackage{amsfonts}
\usepackage{multirow}
\usepackage{amsthm}
\usepackage{float}

\usepackage{comment}
\usepackage{ulem}

\begin{document}

\title{Quantum Simulation of $\mathbb{Z}_2$ Lattice Gauge theory with minimal resources}

\author{Reinis Irmejs}

\affiliation{Max-Planck-Institut für Quantenoptik, Hans-Kopfermann-Str. 1, D-85748 Garching, Germany}
\affiliation{Munich Center for Quantum Science and Technology (MCQST), Schellingstr. 4, D-80799 Munich, Germany}

\author{Mari-Carmen Banuls}
\affiliation{Max-Planck-Institut für Quantenoptik, Hans-Kopfermann-Str. 1, D-85748 Garching, Germany}
\affiliation{Munich Center for Quantum Science and Technology (MCQST), Schellingstr. 4, D-80799 Munich, Germany}

\author{J. Ignacio Cirac}
\affiliation{Max-Planck-Institut für Quantenoptik, Hans-Kopfermann-Str. 1, D-85748 Garching, Germany}
\affiliation{Munich Center for Quantum Science and Technology (MCQST), Schellingstr. 4, D-80799 Munich, Germany}

\date{March 27, 2023}

\begin{abstract}
    The quantum simulation of fermionic gauge field theories is one of the anticipated uses of quantum computers in the NISQ era. 
    Recently work has been done to simulate properties of the fermionic $\mathbb{Z}_2$ gauge field theory in (1+1) D and the pure gauge theory in (2+1) D. 
    In this work, we investigate various options for simulating the fermionic $\mathbb{Z}_2$ gauge field theory in (2+1) D. To simulate the theory on a NISQ device it is vital to minimize both the number of qubits used and the circuit depth. In this work we propose ways to optimize both criteria for simulating time dynamics. In particular, we develop a new way to simulate this theory on a quantum computer, with minimal qubit requirements. We provide a quantum circuit for simulating a single first order Trotter step that minimizes the number of 2-qubit gates needed and gives comparable results to methods requiring more qubits. Furthermore, we investigate variational Trotterization approaches that allow us to further decrease the circuit depth. 
\end{abstract}
\maketitle

\section{Introduction}

Simulating the dynamics of physical quantum systems is one of the most anticipated applications of quantum computing and a good candidate to show useful quantum advantage for a NISQ device \cite{Childs2018}. Physical systems of interest include quantum chemistry models, material simulations, and high energy physics problems via lattice gauge theories, as the one considered here \cite{Bauer2020, Clinton2022, Ba2020}. To simulate the quantum dynamics on a near-term quantum device, the resources used need to be optimized. NISQ devices offer only a limited number of qubits, and have limited coherence times, as well as considerable 2-qubit gate errors \cite{Arute2019}. Thus, to simulate a given problem it is necessary to optimize the number of qubits used and their architecture, as well as the depth (and the number of 2-qubit gates) of the quantum circuit. This is especially true for lattice gauge theories that feature more complicated plaquette and dynamical fermion terms \cite{Zohar2022}.

In this work, we focus on the simulation of the full $\mathbb{Z}_2$ (i.e. including fermionic matter) lattice gauge theory in (2+1) D with minimal resources. In particular, we use as benchmarks the number of qubits and 2-qubit quantum gates needed to implement a single first-order Trotter step. The latter can be used to probe the dynamics of the system either directly, via a Trotterized time evolution, or by using it as a single step for an ansatz to perform variational quantum algorithms for time evolution, like parametrised variational quantum dynamics (pVQD) \cite{Barison2021,  Mansuroglu2021, Cirstoiu}. The same ansatz can also be applied for other algorithms like QAOA or variational quantum eigensolver (VQE) \cite{Peruzzo2014} to probe the ground state properties. In minimizing the resources, we exploit the fermion mapping (fermion elimination method) introduced in \cite{Zohar2018,Zohar2019}, which allows the fermionic $\mathbb{Z}_2$ theory to be encoded with the same number of qubits as the pure gauge theory without the fermions. This is the first practical proposal that evaluates the resources needed for simulating such fermionic (2+1) D physical system on a quantum computer \footnote{ While completing this manuscript, an independent proposal has appeared that explored the use of the same fermion elimination method and also considered the fermionic $\mathbb{Z}_2$ theory in (2+1) D in their work \cite{Greenberg2022}.}.
We compare the circuit depth obtained via the fermion elimination method with the one obtained if a standard approach for encoding fermions - Verstraete-Cirac (VC) \cite{Verstraete2005} transformation is used. The new method offers similar circuit depth requirements , with 17 $CX$ gates per link as compared to the 14 of VC encoding,
 while only using half of the qubits of the latter.  Furthermore, the use of the variational pVQD algorithm is explored to further reduce the requirements for the circuit depth to perform time evolution of the system.

The discretization of continuous gauge field theories on a lattice has enabled very successful numerical results in high energy physics~\cite{Durr2008, Ratti2018b}. 
Here we consider a genuinely discrete lattice gauge theory, namely $\mathbb{Z}_2$  with fermionic matter. This simple model allows an easy encoding of the gauge field in qubits, but merits also an interest of its own. In high energy physics, SU(N) theories are of particular importance, since the strong force, responsible for quark binding and their interactions is mediated via SU(3) gauge field. The exact mechanism of the quark confinement is poorly understood and many insights have been obtained from numerical simulations. In particular, it is believed that the centre of the SU(N) theory - $\mathbb{Z}_N$ is responsible for the confinement \cite{Ikeda2021}. The classical simulations using Monte-Carlo methods suffer from a sign problem and 
 the required resources scale exponentially with the system size. 
 Quantum computers could avoid this problem by working in the Hamiltonian formalism,  
  see \cite{Ba2020, Banuls2020ropp, Zohar2022, USreview} and refs within.
 We show that the circuit depth needed to simulate a single Trotter step is independent of the system size, allowing the simulation to be scaled. As quantum technologies continue to advance, it is important to explore the optimal ways to simulate this theory in a sign problem free way to better understand its properties, and eventually the process of quark confinement. This work only considers the $\mathbb{Z}_2$ theory, but the methods used here can be altered to probe other $\mathbb{Z}_N$ theories, which are left for future work. 

Previous work in \cite{Lumia2022, Gustafson2021} covered the simulation of a pure $\mathbb{Z}_2$ theory in (2+1) D. Very recently, \cite{Mildenberger2022} simulated the fermionic $\mathbb{Z}_2$ theory in (1+1) D, with an implementation on the Google Sycamore quantum device, and particular emphasis on probing the confinement. The authors were able to perform the simulation of time dynamics via Trotterization, with a much greater accuracy than one naively would expect from the 2-qubit gate error rate of the device. Additionally, there have been multiple proposals to simulate the $\mathbb{Z}_2$ theory in both pure and matter case with analog quantum simulators \cite{Zohar2016, Barbiero2019, Homeier2020, Homeier2022, Surace2021b}. These works point out the current interest of simulating the fermionic $\mathbb{Z}_2$ theory in (2+1) D, which we analyze in this paper.

We start by reviewing the Hamiltonian approach to the pure gauge and fermionic $\mathbb{Z}_2$ theory in (2+1) D. Next we introduce the mechanism to encode the fermions in the gauge-field, as proposed in \cite{Zohar2018, Zohar2019}, and how it can be applied to the $\mathbb{Z}_2$ theory. In section~\ref{sec:QC}, we show how this model can be mapped to a quantum circuit and evaluate the necessary number of 2-qubit gates needed for a single step of a first order Trotter circuit, which is compared with the VC method. Section~\ref{sec:numerics} explores the use of the variational methods, including numerical results.
Finally, section~\ref{sec:conclusion} summarizes our conclusions.

\section{$\mathbb{Z}_2$ Lattice Gauge Theory}
\label{sec:Z2lgt}

\begin{figure}
    \centering
    \includegraphics[width = \linewidth]{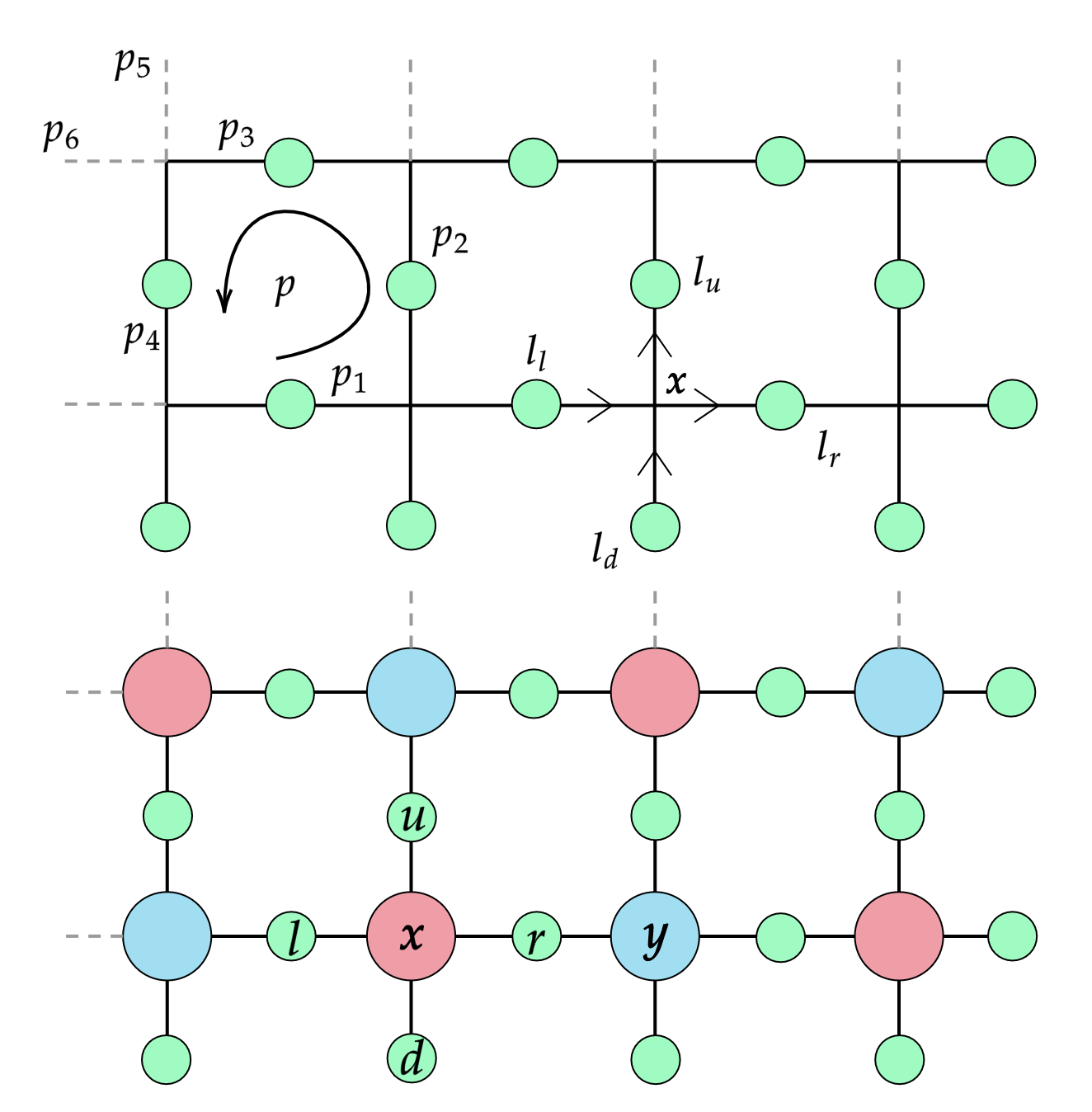}
    \caption{The top figure shows the labelling for the pure gauge theory on a lattice. The bottom figure shows the labelling for the full fermionic theory with staggered fermionic matter. Matter sites are located on vertices while gauge fields are on the links.}
    \label{LatFig}
\end{figure}

In the lattice gauge theory Hamiltonian formalism, the space is discretized, but the time is left continuous. On the lattice, matter fields are located on the vertices (labelled by vectors $\bm{x}$) and the gauge fields on the links (labelled $l$) connecting them. 
In two spatial dimensions, $\bm{x}=(x,y)$. In this case, the fermions can be staggered as shown in Fig~\ref{LatFig}---on even sites (red) we have particles and on the odd sites (blue) antiparticles, with charges $+1$ and $-1$ respectively. The parity of the site is given by $(-1)^{s(\bm{x})} = (-1)^{x+y}$ with $1\; (-1)$ indicating an even (odd) site \cite{Kogut1975}.
The green sites on the links denote the gauge fields. We will consider two-dimensional rectangular lattices with periodic boundary conditions and dimension $M\cross N$, where $M$, $N$ are even, to accommodate fermion staggering. Note that on a lattice of size $M\cross N$, there are $L =2\cross M\cross N$ gauge field links and $M\cross N$ fermion vertices (Fig \ref{LatFig}). 
The gauge field in $\mathbb{Z}_N$ theories has a finite-sized Hilbert space of dimension N, thus allowing it to be encoded on each link.

\subsection{Pure $\mathbb{Z}_2$ lattice gauge theory}

The Hamiltonian of the pure gauge $\mathbb{Z}_2$ theory is given by $H_{KS}$ \cite{Bhanot}:

\begin{align}
    H_{KS} &= H_E + H_B \nonumber \\ &=  \lambda_E \sum_l[2-(P_l + P^\dagger_l)]\\&+ \lambda_B \sum_p[2-(U_{p_1} U_{p_2} U^\dagger_{p_3} U^\dagger_{p_4} + H.c.)].\nonumber
\end{align}
For the electric term $H_E$ the sum is over all links and for the magnetic term $H_B$ over all plaquettes $p$, as indicated in Fig \ref{LatFig}. When discretizing the theory from the continuum one,  the coupling constants are connected through the relations $\lambda_E = g^2/2$, $\lambda_B = 1/2g^2$ \cite{Horn79}. 

For $\mathbb{Z}_2$ the field can take two values: 0 and 1. The theory can be defined using two generators per link $U_l$, $P_l$, satisfying the relations:
\begin{align}
    P_l^2 &= U_l^2 = 1,\\
    P_l^\dagger P_l &= U_l^\dagger U_l = 1,\\
    P_l^\dagger U_l P_l &= \exp(i\pi) U_l = -U_l.
\end{align}
The gauge field on a given link can be encoded into qubit states $\ket{0},\ket{1}$ corresponding to the field values such that $U_l\ket{E} = \ket{(E+1) \mod{2}}$ and $P_l\ket{E} = \exp(i\pi E)\ket{E}$. Thus, $U_l$ is the raising (lowering) operator for the gauge field and $P_l$ is a diagonal operator in this basis, describing the field strength. An additional constraint that the theory obeys is the Gauss law:

\begin{align}
    G(x) &= P_u P_r P_d^\dagger P_l^\dagger \\ &=\nonumber \exp(i\pi[E_u+E_r-E_d-E_l])\\ &= \exp(i\pi Q(\bm{x})) = 1, \nonumber
\end{align}
where $Q(\bm{x})$ is the charge on vertex $\bm{x}$. For the pure gauge theory with no static charges, $Q(\bm{x}) = 0$ on all sites. The sign convention for the gauge fields is given in Fig. \ref{LatFig}.

This Hamiltonian can be implemented on a quantum computer by mapping:
\begin{align}
    U_l &\xrightarrow{} X_l,\\
    P_l &\xrightarrow{} Z_l,
\end{align}
where $(X_l, Y_l, Z_l)$ are Pauli matrices acting on link $l$. The Pure gauge Hamiltonian is expressed as:
\begin{align}
    H_{KS} &= H_E + H_B \nonumber\\ &= -2\lambda_E \sum_l Z_l - 2\lambda_B \sum_p X_{p_1} X_{p_2} X_{p_3} X_{p_4}, 
\end{align}
and the Gauss law (in the absence of external charges) is given by 
\begin{equation}
    Z_u Z_r Z_d Z_l = 1,
\end{equation}
for every vertex.
The quantum circuit to simulate this model will be explicitly shown in the next section.

\subsection{Fermionic $\mathbb{Z}_2$ lattice gauge theory}

In the full theory, when the gauge field interacts with matter, the Hamiltonian acquires two extra terms, the mass term of the dynamical fermions and the interaction term between the fermions and the gauge field. The matter field part of the Hamiltonian is given by:
\begin{align}
    H_{f} &= H_M + H_{int}\nonumber \\&=\sum_x{(-1)^{s(\bm{x})}M a_{\bm{x}}^\dagger a_{\bm{x}}}\nonumber \\&+ \epsilon \sum_{\bm{x}}a_{(x,y)}^\dagger U_r({\bm{x}}) a_{(x+1,y)}+H.c. \nonumber \\&+ \epsilon \sum_{\bm{x}}a_{(x,y)}^\dagger U_u({\bm{x}}) a_{(x,y+1)} +H.c. ,
    \label{eq:Hf}
\end{align}
with $a_i, a_j^\dagger$ satisfying the canonical anticommutation relations (CAR). Note that  in  \eqref{eq:Hf}
the interaction term has been split into horizontal and vertical parts for future convenience. 
To accommodate both particles and antiparticles on the lattice, the staggered fermion approach is used. Here on even (odd) sites we have particles (antiparticles) with charge +1 (-1). 
In this approach, in the computational basis,
\begin{align}\label{meaning}
    &\text{on even site}\begin{cases}\ket{0} \xrightarrow{} \text{ vacuum (no charge)}\\ \ket{1} \xrightarrow{}  \text{ particle (charge +1)} \end{cases}\\ \nonumber
    &\text{on odd site}\begin{cases}\ket{0} \xrightarrow{} \text{ anti-particle (charge -1)}\\ \ket{1} \xrightarrow{}  \text{ vacuum (no charge)} \end{cases}
\end{align}
The number operator on the vertex is given by:
\begin{equation}
    n(\bm{x}) = \frac{1 - (-1)^{s(\bm{x})}Z_x}{2}.
\end{equation}
While the number operator can be easily expressed, the fermionic creation/annihilation operators require some attention, as they obey the non-local CARs. In the method from \cite{Zohar2018,Zohar2019}, that we review in the next section, the fermionic statistics is absorbed into the gauge field, at the expense of increasing the Pauli weight of the Hamiltonian terms (i.e. the number of qubits on which the term acts non-trivially). The transformed Hamiltonian consists of hard-core bosonic matter for which creation/annihilation operators can be implemented with simple spin raising/lowering ones. Furthermore, these hard-core bosonic degrees of freedom can be eliminated by the use of Gauss law, which uniquely determines the charge distribution on the vertices. This allows the full fermionic $\mathbb{Z}_2$ theory to be simulated with the same number of qubits  needed for the pure gauge one, minimizing the spatial resources of the quantum computation. 

\subsection{Fermion encoding via elimination}

In \cite{Zohar2018} a method was introduced to perform a unitary transformation that converts the fermionic degrees of freedom to hard core bosonic degrees of freedom, if the gauge group has $\mathbb{Z}_2$ as a normal subgroup. As a result, the theory acquires phase factors $\xi$ of the gauge field to keep track of the fermionic exchange antisymmetry. The transformed Hamiltonian is
\begin{align}
    H_M &= \sum_x{(-1)^{s(\bm{x})}M\eta_{\bm{x}}^\dagger \eta_{\bm{x}}},\\
    H_{KS} &= - \lambda_E \sum_l(P_l + P^\dagger_l)\nonumber\\& - \lambda_B \sum_p(\xi_p U_{p_1} U_{p_2} U^\dagger_{p_3} U^\dagger_{p_4} + H.c.),\\
    H_I &= \frac{\epsilon}{i} \sum_{\bm{x}}\xi_h({\bm{x}})\eta_{(x,y)}^\dagger U_r({\bm{x}}) \eta_{(x+1,y)}+H.c.\nonumber \\&+\frac{\epsilon}{i}\sum_{\bm{x}} \xi_v(\bm{x})\eta_{(x,y)}^\dagger U_u({\bm{x}}) \eta_{(x,y+1)} +H.c.,
\end{align}
where $\eta({\bm{x}})$ is a staggered hard-core boson annihilation operator and the $\xi$ phase factors are given by: 
\begin{gather}
    \xi_h(x,y) =  (-1)^{E_u(x,y)+E_l(x,y)+E_d(x,y)_+E_d(x+1,y)},\\
    \xi_v(x,y)= (-1)^{E_l(x,y)+E_d(x,y)},\\
    \xi_p = (-1)^{E_{p_1}+E_{p_2}+E_{p_5}+E_{p_6}},
\end{gather}
and the ordering is shown in Fig \ref{LatFig}. Under this transformation, the Gauss law remains unchanged. This is important, as the Gauss law fully defines the charge configuration on the vertex and thus can be used to eliminate the matter fields \cite{Zohar2019}. 

For an occupied (unoccupied) site ($n(x) = 1 (0)$) $Q(x) = \pm 1 (0)$ and the Gauss law gives $P_u P_r P_d^\dagger P_l^\dagger = -1\; (1)$. We define projection operators $\Pi_{\rho}(x,y)$ that project the Hilbert space to the subspace with $G(x) = \rho$, with $\rho = 1$ indicating that the site is empty and $\rho = -1$ that the site is full. Elimination of the matter fields via Gauss law leads to the $H_f$ terms to acquire projection operators as follows:
\begin{align}
    H_f &= H_M + H_I\nonumber =\sum_{\bm{x}}M \Pi_{-1}(\bm{x})+\nonumber\\-i \epsilon &\left( \sum_{\bm{x}}(-1)^{s(\bm{x})}\xi_h(\bm{x})\Pi_{-1}(x,y) U_r(x,y) \Pi_{1}(x+1,y) +\right. \nonumber\\ & \left. \sum_{\bm{x}}(-1)^{s(\bm{x})}\xi_v(\bm{x})\Pi_{-1}(x,y) U_u(x,y) \Pi_{1}(x,y+1) \right) \nonumber \\ &+ h.c.,
\end{align}
where the factors of $(-1)^{s(\bm{x})}$ arise from fermion staggering.
Thus, the full fermionic $\mathbb{Z}_2$ theory can be simulated only by encoding the gauge field values. Once again, it is worth re-emphasizing that the matter fields have been eliminated at the expense of the Gauss law, thus leaving the new theory without this constraint.  In further sections it will be shown how each of these terms can be encoded on a digital quantum computer.  

\subsection{Full $\mathbb{Z}_2$ theory as a spin system}

Here we show how the pure gauge Hamiltonian can be simulated on a quantum computer. We will assume access to Pauli gates $\mathcal{P} = \{X,Y,Z\}$ and their single qubit rotations $\{RX(\theta), RY(\theta), RZ(\theta)\}$ with $R\mathcal{P}(\theta) = \exp(-i\theta\mathcal{P}/2)$
as well as controlled $X$ ($CX$) gate as the 2 qubit gate. To implement this model, it is necessary to have an architecture of qubits with a possibility to perform the $CX$ gate between nearest neighbours.

In order to perform the simulation, in addition to the mapping of $U_l \xrightarrow{} X_l$ and $P_l \xrightarrow{} Z_l$ introduced previously, we need to map the projection operators $\Pi_g$ and the phase factors $\xi$. The mapping of the phase factors is straight forward since $(-1)^{E_l} = P_l$. The projection operator $\Pi_\rho$ can be implemented as follows:
\begin{equation}
    \Pi_{\pm 1} (x,y) = \frac{1}{2} (1 \pm Z_u Z_l Z_d Z_r (x,y)) = \frac{1}{2} (1 \pm G(x,y)).
\end{equation}
The Hamiltonian mass term $H_M$ is thus given by:
\begin{equation}
    H_M = \sum_x{\frac{M}{2}(1-Z_{u}Z_{r}Z_{d}Z_{l})}.\\
\end{equation}

Since we know how to implement each operator in the interaction Hamiltonian $H_{int}$, it can be mapped to a quantum device. While the direct mapping produces a somehow complicated structure, it can be considerably simplified as follows.
Consider the horizontal part of the interaction Hamiltonian and note that $U_l = U_l^\dagger = X_l$.

\begin{widetext}
\begin{align}
    &H_{H} = \nonumber\\ &=\frac{\epsilon}{i} \sum_{\bm{x}}(-1)^{s(\bm{x})}\xi_h(\bm{x}){\Pi_{-1}(x,y) U_r(x,y) \Pi_{1}(x+1,y)} +h.c. = \frac{\epsilon}{i}\sum_{\bm{x}}(-1)^{s(\bm{x})}U_r(x,y)\xi_h(\bm{x})\Pi_{1}(x,y) \Pi_{1}(x+1,y)+h.c.\nonumber\\
    &= \frac{\epsilon}{i}\sum_{\bm{x}} (-1)^{s(\bm{x})} U_r(x,y)\xi_h(\bm{x})(Z_r (x,y))^2\Pi_{1}(x,y)\Pi_{1}(x+1,y) +h.c.\nonumber\\
    &= \frac{\epsilon}{i}\sum_{\bm{x}} (-1)^{s(\bm{x})} U_r(x,y)G(\bm{x})Z_d(x+1,y)Z_r (x,y)\Pi_{1}(x,y) \Pi_{1}(x+1,y)+h.c.\nonumber\\
    &= -\epsilon\sum_{\bm{x}} (-1)^{s(\bm{x})} Y_r(x,y)Z_d(x+1,y){\Pi_{1}(x,y) \Pi_{1}(x+1,y)}+h.c.\nonumber\\
    &= -\epsilon\sum_{\bm{x}} (-1)^{s(\bm{x})} Y_r(x,y)Z_d(x+1,y){\Pi_{1}(x,y) \Pi_{1}(x+1,y)}-\epsilon\sum_{\bm{x}} (-1)^{s(\bm{x})} Y_r(x,y)Z_d(x+1,y){\Pi_{0}(x,y) \Pi_{0}(x+1,y)}\nonumber\\ 
    &= -\frac{\epsilon}{2}\sum_{\bm{x}} (-1)^{s(\bm{x})} Y_r(x,y)Z_d(x+1,y)(1+Z_u Z_l Z_d(x,y)\cross Z_d Z_r Z_u(x+1,y)).
\end{align}
\end{widetext}
In the lines 1-3, the projection operators are collected, followed by insertion of $(Z_r (x,y))^2$ and simplification from the Gauss law constraint in line 4. In the last 2 lines, both terms are collected and $\Pi_{\rho}$ values are inserted to give the final result. 

Similarly the vertical interaction terms can be simplified to:
\begin{align}
    &H_{V} = 
    -\epsilon\sum_{\bm{x}} (-1)^{s(\bm{x})} Y_u(x,y)Z_r(x,y)\nonumber\\&\cross \frac{1}{2}(1+Z_r Z_l Z_d(x,y)\cross Z_l Z_r Z_u(x,y+1)).
\end{align}
The interpretation of these terms is that we will have an interaction term of the form $Y\otimes Z$ acting when both sites at the end of the links are empty or occupied. This corresponds to either particle-antiparticle pair creation or annihilation. The pure gauge part of the Hamiltonian gets slightly altered, with the plaquette term becoming 6-local:
\begin{align}
    H_{KS} &= H_E + H_B \nonumber\\ &= -2\lambda_E \sum_l Z_l - 2\lambda_B \sum_p Y_{p_1} Y_{p_2} X_{p_3} X_{p_4} Z_{p_5} Z_{p_6}.
\end{align}

In this matter-eliminated formalism, the most complicated terms to implement are the interaction and magnetic ones since they are both 6-local. Despite the complications introduced by the projectors, the final gate complexity to perform time evolution is similar to using the Verstraete-Cirac encoding.

The same model was also considered in \cite{Greenberg2022} where the authors arrived at the same result. A similar (transformed) Hamiltonian was obtained in \cite{Moroz2022} for classical simulation. In our work the emphasis is put towards optimization for circuit depth and comparison with other methods.

\subsection{Fermion encoding via VC transformation}

Different methods exist to deal with the fermion statistics in simulations. 
The simplest strategy is to encode the fermions via Jordan-Wigner transformation \cite{JordanWigner},
effective for one-dimensional (or small two-dimensional) systems. In this transformation, fermions in a chosen order are mapped to spin operators, keeping track of their CARs. However, in two dimensions, any such ordering maps local fermionic terms (e.g. nearest-neighbor interactions) to non-local ones, which results in strings of spin operators.
 In general, the Pauli weight of interaction terms after this mapping will scale as $\mathcal{O}(L)$ where $L$ is the linear size of the 2D system.

There exist several fermionic encoding methods that map a local fermion Hamiltonian to a local spin system \cite{Verstraete2005,Steudtner2019,Derby2020}.
However, in all of these methods extra spins (qubits) are introduced to enforce the fermion CARs, thus making them unfavourable in terms of the spacial quantum computation requirements when compared to the fermion elimination method.
One such method - the Verstraete-Cirac (VC) transformation \cite{Verstraete2005} encodes fermions as spins by introducing ancillary qubits and encoding the fermionic statistics into this multi-qubit increased Hilbert space. Despite the fact that this method has been around for nearly two decades it is still one of the lowest-weight encodings, and a gold standard for benchmarking. In the VC approach, the pure gauge part of the Hamiltonian remains unchanged, with the matter part of Hamiltonian increasing in weight. To accommodate for fermion statistics, an extra qubit is introduced per each fermion site and the operators acting on the ancillary qubits are denoted by $\Tilde{A}$. Under this mapping, the matter Hamiltonian becomes:
\begin{align}
    H_{f} &= H_M + H_{int} \nonumber\\&=\sum_x{\frac{(-1)^{s(\bm{x})}M}{2} Z(\bm{x})}+\nonumber\\& \sum_{\bm{x}} \epsilon_h(\bm{x})X_r(\bm{x}) (X(x,y) X(x+1,y)\nonumber \\&+ Y(x,y) Y(x+1,y))\Tilde{Z}(\bm{x})\nonumber \\&+ \sum_{\bm{x}}\epsilon_v(\bm{x})X_u(\bm{x})(X \Tilde{Y}(x,y) Y \Tilde{X}(x,y+1)\nonumber\\&- Y \Tilde{Y}(x,y) X \Tilde{X}(x,y+1) ).
\end{align}
Each of the horizontal terms has 2 components, each of them with weight 4, while the vertical components have weight 5. 

\section{Quantum Circuit Methods}
\label{sec:QC}

\subsection{Simulating time dynamics via Trotterization}

Trotterization is a common way of simulating time dynamics in which the non-local exponential of a Hamiltonian is approximated as a sequence of smaller, easier to implement unitaries, by means of a Suzuki-Trotter expansion. In this method the entire time evolution gets divided into $n = t/\delta$ steps of fixed size $\delta$, as
\begin{equation}
    U(t) = \exp(-iHt) = (\exp(-iH\delta))^{t/\delta}.
\end{equation}
In general the Hamiltonian $H$ contains multiple terms that do not commute. Such Hamiltonian can be written as $H = \sum_{i=1}^\mathcal{M} H_i$ where each $H_i$ does not commute with the others, but all terms within each of them do. At the lowest Trotter order, each step $U(\delta)$ is approximated as $\mathcal{V}(\delta)$:
\begin{equation}
    U(\delta)\simeq \mathcal{V}(\delta) = \prod_{i=1}^\mathcal{M}\exp(-i\delta H_i).
\end{equation}
In general a single Trotter step of a Hamiltonian, composed as a sum of local Pauli terms can be expressed as:

\begin{equation}
    \mathcal{V}(\delta) = \prod_{i=\mathcal{M}}^{1} \prod_{\mathcal{L}} (V_i R\mathcal{P}_i(c (\delta))V_0,
\end{equation}
where $V_i$ is a general unitary operator and $R\mathcal{P}(c(\delta))$ is a Pauli operator rotation that depends on the time step size $\delta$ and the Hamiltonian couplings. Note that we have used the fact that in each $H_i$ the $\mathcal{L}$ commuting terms can be done in parallel. Next, the exact form of each of the terms $R\mathcal{P}_i, V_i$ will be given.

The entire error for the time evolution with Trotterization can be bound by \cite{Childs2021}:
\begin{equation*}
    \norm{U(t)-\mathcal{V}(t)} \le \frac{t\delta}{2}\sum_{i = 1}^{\mathcal{M}}\norm{\sum_{j = i+1}^\mathcal{M} \comm{H_i}{H_j}},
\end{equation*}
and thus it depends on the time step $\delta$ and the total evolution time $t$. 
Furthermore, it has been observed that in practice these bounds are loose and the Trotter error tends to be much smaller  \cite{Childs2021}. Recent studies \cite{Heyl2019a, Chinni2021} that explored the chaos-regular transition in Trotterized quantum dynamics showed that even for large values of $\delta$ the systems still obey controlled behavior. Furthermore, the threshold for this transition is largely independent of the system size considered. This is an important result as it illustrates that one can faithfully probe time dynamics with large $\delta$ values, thus minimizing the number of steps and the circuit depth needed to perform a simulation of a given time. 

\subsection{Quantum Circuit for pure gauge $\mathbb{Z}_2$}
\label{sec:QC_KS}

In the pure $\mathbb{Z}_2$ case there are two non-commuting parts $H_E$ and $H_B$. To perform time evolution we need to implement both $\exp(-i \delta H_E)$ and $\exp(-i\delta H_B)$. Implementing the $\exp(-i \theta H_E)$ is trivial in the chosen basis as it is just a $RZ$ rotation on each link:

\begin{figure}[H]
\centering
\begin{quantikz}
\lstick{} & \gate{RZ(2\theta)} & \qw \\
\end{quantikz}
\end{figure}

Implementing the terms in $H_B$ is more difficult. Note that a weight $K$ Pauli term can be implemented with $(2K-2)$ $CX$ gates. For the terms appearing in $H_B$ of form $\exp(i\theta X^{\otimes 4})$, the identity $X_a X_b = CX_{ab} X_a CX_{ab}$ can be used to yield:
\begin{figure}[H]
\centering
\begin{quantikz}
\lstick{$p_1$} & \ctrl{1} & \ctrl{2} & \gate{RX(2\theta)}&\ctrl{2}&\ctrl{1}&\qw\\
\lstick{$p_2$} & \targ{1}  & \qw &\qw &\qw &\targ{1}&\qw\\
\lstick{$p_3$} & \ctrl{1}  & \targ{}&\qw   &\targ{} & \ctrl{1}&\qw\\
\lstick{$p_4$} & \targ{1} & \qw &\qw& \qw&\targ{1}&\qw\\
\end{quantikz}
\end{figure}
Thus, a single Trotter step of the pure theory can be implemented with $6\cross 1/2 \cross L = (3\cross L)$ 2-qubit $CX$ gates for a system with $L$ links.

\subsection{Quantum Circuit for fermionic $\mathbb{Z}_2$}

In the fermionic $\mathbb{Z}_2$ theory, we need to implement all 5 terms - $H_E$, $H_B$, $H_M$, $H_H$, $H_V$.
\begin{enumerate}
    \item The implementation of the $\exp(-i\delta H_E)$ is the same as in the pure case, it consists of $\exp(-i\theta Z)$ rotations that can be done in parallel on each link:
    \begin{figure}[H]
    \centering
        \begin{quantikz}
            \lstick{} & \gate{RZ(2\theta)} & \qw 
        \end{quantikz}
    \end{figure}
    \item The implementation of $\exp(-i\delta H_B)$ is slightly more complicated than in the pure case as it is 6-local. Each term is of the form $\exp(-i\theta Y_{p_1} Y_{p_2} X_{p_3} X_{p_4} Z_{p_5} Z_{p_6} )$ and can be implemented as $V_2^\dagger RX_{p_3}(2\theta) V_2$ where the circuit $V_2$ is given by:
    \begin{figure}[H]
        \centering
        \begin{quantikz}
            \lstick{$p_1$} & \gate{RZ(-\frac{\pi}{2})} & \gate{H}&\ctrl{1}&\qw &\qw \\
            \lstick{$p_2$}& \gate{RZ(-\frac{\pi}{2})}&\gate{H}&\targ{1}&\ctrl{1}&\qw \\ \lstick{$p_3$}&\gate{H}&\targ{}&\targ{}&\targ{}&\qw \\
            \lstick{$p_4$}&\gate{H}&\ctrl{-1}&\qw&\qw&\qw \\
            \lstick{$p_5$}&\targ{}&\qw&\ctrl{-2}&\qw&\qw \\
            \lstick{$p_6$}&\ctrl{-1}&\qw&\qw&\qw&\qw 
        \end{quantikz}
    \end{figure}
    \item The evolution of the mass term $\exp(-i\delta H_M)$ where each term is of form $\exp(-i\theta Z^{\otimes 4})$ and can be implemented as $\exp(-i\theta Z^{\otimes 4}) = V_3^\dagger RZ_r(2\theta) V_3$ where $V_3$ is given by:
    \begin{figure}[H]
    \centering
    \begin{quantikz}
    \lstick{u} & \targ{1} & \ctrl{3} &\qw \\
    \lstick{l} & \ctrl{-1}  & \qw&\qw \\
    \lstick{d} & \ctrl{1}&\qw&\qw \\
    \lstick{r} & \targ{} & \targ{} & \qw
    \end{quantikz}
    \end{figure}
    The $V_3$ circuit can be interpreted as calculating the parity of a given vertex on link $r$.
    \item The weight-6 part of the horizontal interaction term evolution  $\exp(-i\delta H_H)$ can be implemented as $V_4^\dagger RZ_{L}(2\theta)V_4$ with $V_4$ given by:
    \begin{figure}[H]
    \centering
    \begin{quantikz}
    \lstick{u, (x,y)} & \targ{} &\qw& \ctrl{3}\qw &\qw&\qw&\qw\\
    \lstick{l, (x,y)} & \ctrl{-1}  & \qw &\qw&\qw&\qw&\qw\\
    \lstick{d, (x,y)} & \qw & \qw&\qw&\ctrl{1}&\qw&\qw\\
    \lstick{L = link}  &\gate{RZ(-\frac{\pi}{2})}& \gate{H}&\targ{} & \targ{}&\targ{}&\qw\\
    \lstick{u, (x+1,y)}  & \targ{1} & \qw&\qw&\qw&\ctrl{-1}&\qw\\
    \lstick{r, (x+1, y)}  & \ctrl{-1}  & \qw &\qw&\qw&\qw&\qw\\
    \end{quantikz}
    \end{figure}    
    
    \item The weight-6 part of the vertical interaction term $\exp(-i\delta H_V)$ is similar to the horizontal, making the structure of the circuit similar. Each term can be implemented as $V_5^\dagger RZ_{L}(2\theta)V_5$ with $V_5$ given by:
    \begin{figure}[H]
    \centering
    \begin{quantikz}
    \lstick{u, (x,y)} & \targ{} &\ctrl{2}&\qw&\qw&\qw\\
    \lstick{l, (x,y)} & \ctrl{-1}  & \qw &\qw&\qw&\qw\\
    \lstick{r, (x,y)} & \qw & \targ{}&\ctrl{1}&\qw&\qw\\
    \lstick{L = link}  &\gate{RZ(-\frac{\pi}{2})}& \gate{H} & \targ{}&\targ{}&\qw\\
    \lstick{l, (x,y+1)}  & \targ{1} &\qw&\qw&\ctrl{-1}&\qw\\
    \lstick{d, (x, y+1)}  & \ctrl{-1}  & \qw &\qw&\qw&\qw\\
    \end{quantikz}
    \end{figure}
\item Both the horizontal and vertical terms also have a weight-2 component that can be implemented as $\exp(-i\theta Z_a Y_b) = V^{\dagger}_6 RZ_b(2\theta)V_6$ with $V_6$ given by:
\begin{figure}[H]
\centering
\begin{quantikz}
\lstick{a} &\qw&\qw&\ctrl{1}&\qw\\
\lstick{b}&\gate{RZ(-\frac{\pi}{2})}& \gate{H}&\targ{}&\qw\\
\end{quantikz}
\end{figure}
\end{enumerate}

Note that in all of the circuits the control gates act only between qubits that are nearest neighbours on the lattice. 

If the ordering of the terms is chosen in an optimal way, it is possible to simplify the unitaries by contracting some of the $CX$ gates into identities. Trivially, to apply all these terms one would need $10 \cross 1/2L+6 \cross 1/2 L + 12\cross L = 20 L$ of $CX$ gates (Table \ref{tableCost}). By choosing this order optimally it can be brought down to $17L$ of $CX$ gates for $L$ links. A detailed description of the optimal ordering to obtain this simplified result is given in the Appendix \ref{Ordering}. 
\begin{table}
\begin{center}
    \begin{tabular}{c c c c}
    \hline
    \hline
    The $H$ term & Number of terms & Single cost & Total cost\\
    \hline
    $H_E$ & $L$ & 0 & 0\\
    $H_B$ & $L$/2 & 10 & 5$L$\\
    $H_M$ & $L$/2 & 6 & 3$L$ \\
    $H_H$ & $L$/2 & 12 & 6$L$\\
    $H_V$ & $L$/2 & 12 & 6$L$ \\
    \hline
    Total & & & 20$L$ \\
    Total Reduced & & & 17$L$\\
    \hline
    \hline
    \end{tabular}
    \caption{Table shows the cost of implementing each term of the Hamiltonian in terms of $CX$ gates.}
    \label{tableCost}
\end{center}
\end{table}

Even though this fermion-eliminated Hamiltonian has a complicated structure, the necessary number of 2-qubit gates is quite modest. In comparison, the VC method needs $14L$ $CX$ gates, but it achieves that by using twice as many qubits. 

\subsection{Approaches for circuit depth minimization}
\label{subsec:pVQD}

One possible way to decrease the circuit depth of a particular time dynamics simulation is to use variational methods, such as parametrized variational quantum dynamics (pVQD)~\cite{Barison2021}. The variational approaches allow one to decrease the circuit depth at the expense of executing the quantum circuit multiple times in the optimization subroutine. 

In the pure $\mathbb{Z}_2$ theory, a Trotter step is given by
\begin{equation}
    U(\delta) = \exp(-iH_B \delta)\exp(-iH_E \delta).
\end{equation}
The full time evolution for time $t$ can either be implemented by applying $n$ Trotter steps such that $n\delta = t$, or approximated by a variational circuit. A good candidate for the variational circuit is to simply take $k$ variational steps and optimize the evolution parameters $\theta_i$:

\begin{equation}
    U_{var}(\vec{\theta}) = \prod_{j = 1}^k (\exp(-iH_B \theta_{2j})\exp(-iH_E \theta_{2j+1})).
\end{equation}

The optimization proceeds as follows:

\begin{enumerate}
    \item Start with an easily preparable state $\ket{\Psi}$ to be evolved
    \item For the first step maximize the overlap
    \begin{equation}
        \bra{\Psi}U^\dagger (\vec{\theta}^{1})U(\delta)\ket{\Psi}.
    \end{equation}
    \item Proceed to variationally maximize the overlap:
    \begin{equation}
        C(\vec{\theta}) = \bra{\Psi}U^\dagger(\vec{\theta}^{(i-1)})U(\delta)U(\vec{\theta}^{(i)})\ket{\Psi},
    \end{equation}
    by changing the parameters ${\vec{\theta}^{(i)}}$ and using the already optimized parameters ${\vec{\theta}^{(i-1)}}$ from the previous timestep.
\end{enumerate}
By saving the variational parameters $\vec{\theta}^{(i)}$, it is possible to implement the entire time evolution with the constant circuit depth of $2k+1$ (Trotter timesteps).

For the full fermionic theory, the ansatz can be constructed in a similar way:
\begin{align}\label{ansatz}
    U_{var}(\vec{\theta}) &= \prod_{j =1}^{k}(\exp(-i\theta_{5j}H_B)\exp(-i\theta_{5j+1}H_E))\nonumber\\&\exp(-i(-1)^{x+y}\theta_{5j+2}H_V)\exp(-i\theta_{5j+3}H_M)\\&\exp(-i(-1)^{x+y}\theta_{5j+4}H_H)).\nonumber
\end{align}
Even though here we only explore the application of the ansatz for simulating the time dynamics, it can also be used in variational algorithms like QAOA and VQE to probe the ground state properties of the system. 

\section{Numerical results}
\label{sec:numerics}

In this section we present the numerical results obtained using pVQD. Probing the time dynamics via Trotterization requires to repeat the single Trotter step circuit many times, which results in a large circuit depth and thus makes it hard to execute such simulations on NISQ hardware. But the depth can be kept small and constant with the use of variational methods. Here we apply pVQD to both the pure gauge and the fermionic $\mathbb{Z}_2$ theories, focusing on a $2\cross 2$ lattice. We will explore how the accuracy depends on the number of Trotter layers used in the ansatz \eqref{ansatz}. In all simulations we ignore the shot noise that arises from finite amount of measurements. We consider the evolution from the initial product state:
\begin{equation}
    \ket{\Psi_0} = \prod_{1}^L \ket{0}.
\end{equation}
and measure the accuracy of the variational evolution by the fidelity of the pVQD approximation $\mathcal{F}$,
\begin{equation}
    \mathcal{F}(t) = \abs{\bra{\Psi_0} \mathcal{V}^\dagger(t)U_{var}(\theta)\ket{\Psi_0}}^2.
\end{equation}
The error of the approximation is $1-\mathcal{F}(t)$.

\subsection{Pure Gauge results}
\begin{figure}
    \centering
    {\includegraphics[width=\linewidth]{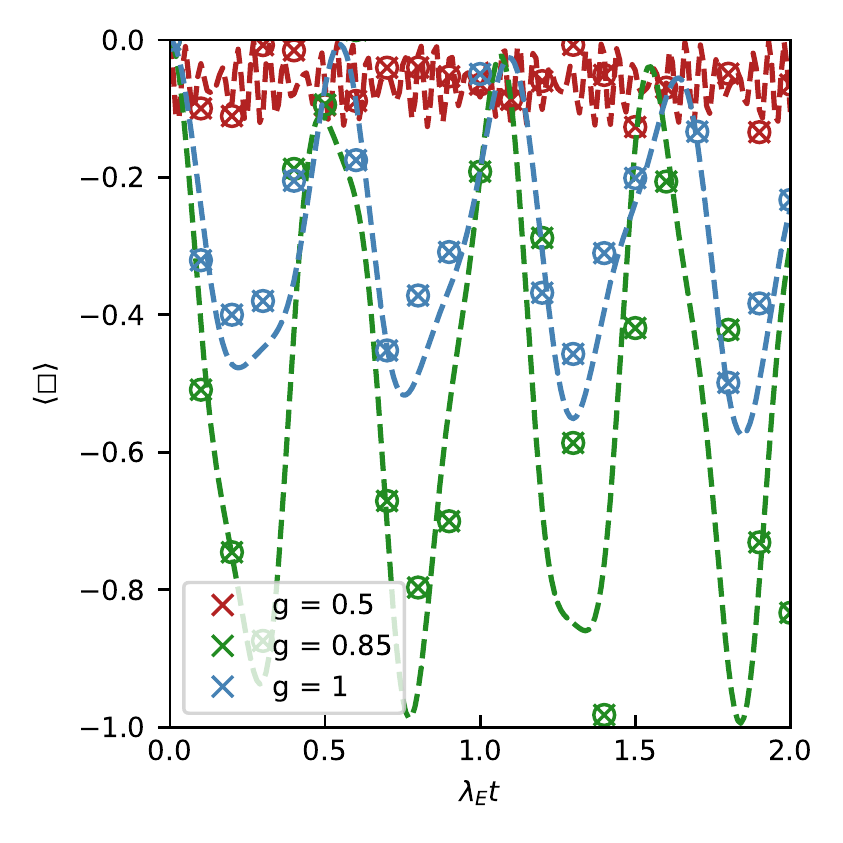}}
    \caption{Figure shows the pVQD results for $g = 0.5,\, 0.85,\, 1$ (lines red, green, blue) of a pure $\mathbb{Z}_2$ on a $2\cross 2$ lattice. In this case the depth of $k = 2$ was used. $\cross$ marks the pVQD results, $\circ$ - Trotterization results, and the dotted lines represent results obtained from exact diagonalization.}
    \label{FigPureG}
\end{figure}

For the pure $\mathbb{Z}_2$ theory we investigate the $2\cross 2$ lattice and observe that the ansatz of $k=2$ steps already well approximate the dynamics for all coupling values considered $g = 0.5, 0.85, 1$. The results are shown in Fig~\ref{FigPureG}. In particular, we look at the expectation value of the plaquette operator $\expval{\square}$ on site $\bm{x} = (0,0)$. The results show excellent agreement for the entire evolution range.
By using this method, it is possible to reduce circuit depth required from 20 to 5 Trotter steps that were used in the pVQD optimization procedure.

\subsection{Full fermionic results}

\begin{figure}
    \includegraphics{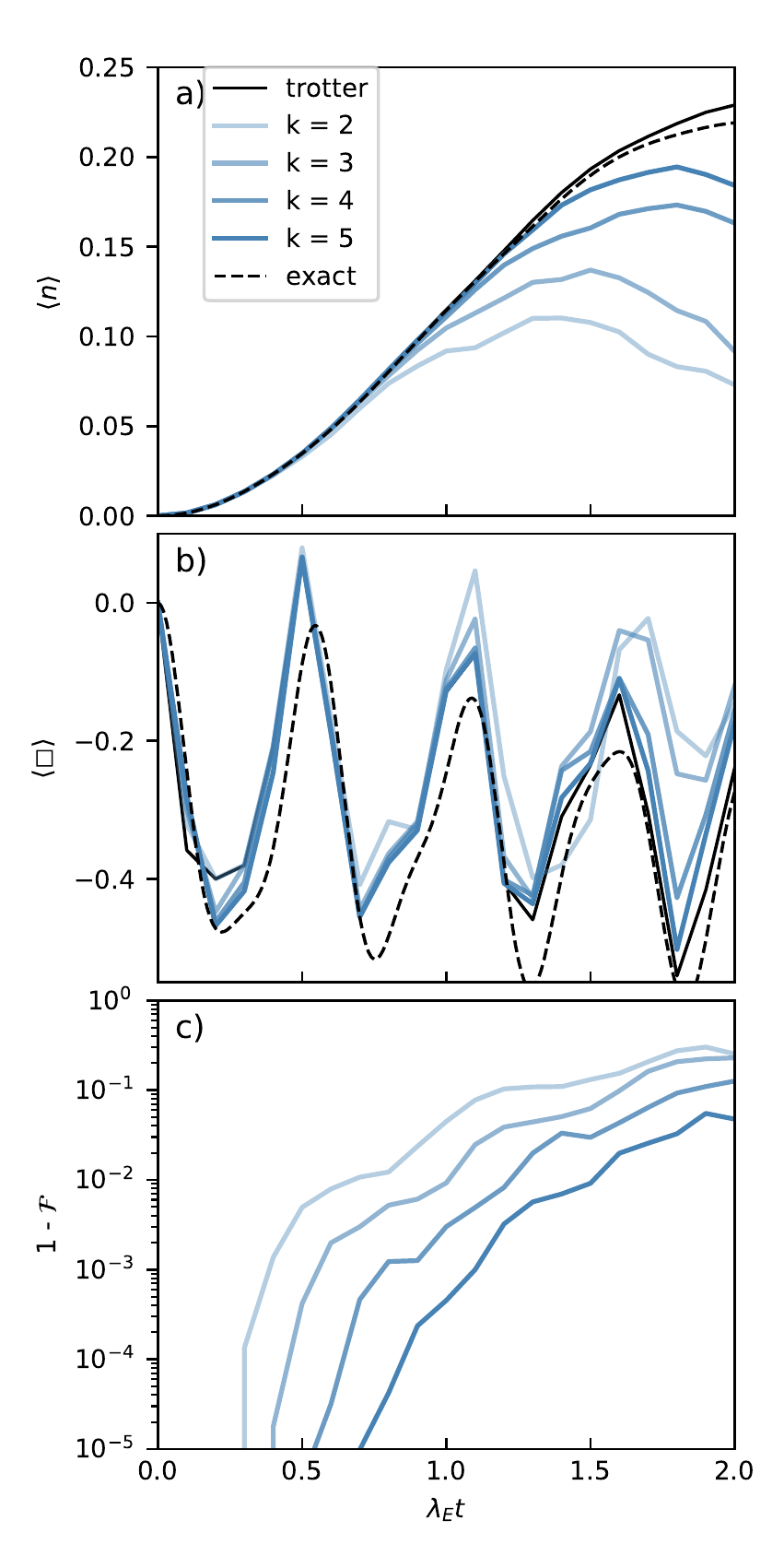}
    \caption{Figure shows the expressibility for the pVQD variational ansatz for depths $k = 2,3,4,5$ for the $2 \cross 2$ fermionic Hamiltonian with couplings $(\lambda_E,\lambda_B,\epsilon,M) = (1,1,0.2,1)$. The time step for the evolution is $\lambda_E\delta = 0.1$. a) shows the expectation value for the occupation $\expval{n}$ on the site (0,0) for various ansatz depths and the comparison with the Trotterized and exact values. b) shows the expectation value of the transformed plaquette operator and the comparison with the Trotterized and exact values. c) shows the accuracy of the approximation in terms of infidelity $1-\mathcal{F}(t) = 1-\abs{\bra{\Psi_{Trotter}(t)}\ket{\Psi_{pVQD}(t)}}^2$.}
    \label{FigFullZ2}
\end{figure}

The variational ansatz for the full fermionic theory consists of 5 terms compared to the 2 for the pure case. This leads to the optimization process being slower and makes it more difficult to reach the global minimum. In this case, we compare the results for depth values of $k = 2,3,4,5$ along with their associated errors for the Hamiltonian with $(\lambda_E,\lambda_B,\epsilon,M) = (1,0.2,1)$ and observe good agreement with the Trotterized evolution (Fig~\ref{FigFullZ2}). We investigate the expectation value of occupation $\expval{n}$ on site $\bm{x} = (0,0)$ and the expectation value of a plaquette operator $\expval{\square}$ on site $\bm{x} = (0,0)$ when evolved under the variational circuit. Furthermore, we study the infidelity $1-\mathcal{F}$ of the variational state when compared to the Trotterized evolution. The results are also compared with the exact dynamics obtained by exact diagonalization. As expected, the increase in the variational circuit depth leads to a better agreement for long times. 

\section{Conclusion}
\label{sec:conclusion}

Here we have presented a new method to simulate the full fermionic $\mathbb{Z}_2$ theory in (2+1) D with minimal resources, in particular, with minimal number of qubits. This was achieved by eliminating the fermionic degrees of freedom and absorbing them into the gauge field. For a lattice of size $M \cross N$ one needs $L =2\cross M \cross N$ qubits (i.e. one per link) to simulate the model. 
In methods that involve encoding the fermions with the help of ancillary qubits, like the Verstraete-Cirac encoding, one needs twice as many qubits. We have shown that the circuit depth in our case is only slightly worse, with 17 $CX$ gates per link, compared to the 14 of VC. Furthermore, we have presented a variational Trotterization strategy that allows to further decrease the circuit depth.
Numerical results of the $2\times 2$ lattice simulation suggest that the time dynamics of both the pure gauge and fermionic $\mathbb{Z}_2$ theory can be well approximated with Trotterized time dynamics. 
For the pure gauge theory, the long time dynamics could be approximated well with a variational ansatz of only $k=2$ layers. For the full fermionic case, the evolution can still be approximated by the variational ansatz, but we find that the number of variational layers need to be increased to obtain reliable results for longer times. 
This work shows that the fermion elimination method is an optimal approach for simulating the $\mathbb{Z}_2$ theory on a quantum computer, due to its minimal qubit requirement and the comparable 2-qubit gate count with other methods. 
Further work involves developing similar methods for higher $\mathbb{Z}_N$ theories and extending them to (3+1) D. \\[0.7in]
While completing this manuscript, an independent proposal appeared that also explores the use of fermion elimination method for simulating  lattice gauge theories, including the fermionic $\mathbb{Z}_2$ in (2+1) D \cite{Greenberg2022}.

\acknowledgements
This work was partially funded by the Deutsche Forschungsgemeinschaft (DFG, German Research Foundation) under Germany's Excellence Strategy -- EXC-2111 -- 390814868 and by the European Union through the ERC grant QUENOCOBA, ERC-2016-ADG (Grant no. 742102), and by the German Federal Ministry of Education
and Research (BMBF) through EQUAHUMO (Grant
No. 13N16066) within the funding program quantum
technologies - from basic research to market and by the
Munich Quantum Valley (MQV), which is supported by
the Bavarian state government with funds from the Hightech
Agenda Bayern Plus, and by the EU-QUANTERA project TNiSQ (BA 6059/1-1).

\appendix

\section{Quantum Circuit using the Verstraete Cirac encoding}

The VC encoded Hamiltonian has $2L$ qubits ($L$ qubits for the gauge field, $L/2$ qubits for matter sites and $L/2$ qubtis for extra ancillas). The VC Hamiltonian is $H = H_{KS}+H_f$. The terms in the pure gauge Hamiltonian $H_{KS}$ can be simulated with $3L$ $CX$ gates per link (section~\ref{sec:QC_KS}). When mapped to qubits using VC transformation the $H_f$ term is given by:
\begin{align}
    H_{f} &= H_M + H_{int} \nonumber\\&=\sum_x{\frac{(-1)^{s(\bm{x})}M}{2} Z(\bm{x})}+\nonumber\\& \sum_{\bm{x}} \epsilon_h(\bm{x})X_r(\bm{x}) (X(x,y) X(x+1,y) \\&+ Y(x,y) Y(x+1,y))\Tilde{Z}(\bm{x})\nonumber \\&+ \sum_{\bm{x}}\epsilon_v(\bm{x})X_u(\bm{x})(X \Tilde{Y}(x,y) Y \Tilde{X}(x,y+1)\nonumber\\&- Y \Tilde{Y}(x,y) X \Tilde{X}(x,y+1) ).\nonumber
\end{align}
The mass term $H_M$ can be implemented trivially since it is only an $RZ$ gate.
The horizontal interaction term for each link has 2 weight-4 terms. For each of the terms we want to implement a rotation of type:
\begin{equation}
    \exp(i\theta X_1 X_2 X_3 Z_4)\exp(i\theta X_1 Y_2 Y_3 Z_4).
\end{equation}
Again, by using similar methods as before, this can be decomposed as:
\begin{align}
    &\exp(i\theta X_1 X_2 X_3 Z_4)\exp(i\theta X_1 Y_2 Y_3 Z_4)\nonumber\\
    &= H_4 \exp(i\theta X_1 X_2 X_3 X_4) RZ_2(-\pi/2)RZ_3(-\pi/2)\\
    &\cross\exp(i\theta X_1 X_2 X_3 X_4)RZ_2(\pi/2)RZ_3(\pi/2) H_4,\nonumber
\end{align}
which can be implemented with 10 $CX$ gates.
Similarly, the vertical interaction terms can be implemented with 12 $CX$ gates. 

Thus, the total cost for implementing a single step of Trotterized time evolution is $(3L + (10+12)\cross L/2 = 14L$ $CX$ gates.

\section{Ordering of the terms}\label{Ordering}

The optimization for the $CX$ gate count comes from picking the optimal order in which to implement each term in the Trotterized evolution:
\begin{itemize}
    \item Start with Implementing $H_E$ on all vertices, followed by the weight-2 part of the $H_H$ and $H_V$ terms.
    \item Then start with (even,even) vertices. Perform the weight-6 part of $H_H$, followed by $H_M$, followed by $H_B$, followed by the weight-6 part of $H_V$.
    \item Next, starting from (even,odd). Perform the weight-6 part of $H_H$, followed by $H_M$, followed by $H_B$, followed by the weight-6 part of $H_V$.
    \item Next, starting from (odd,even). Perform the weight-6 part of $H_H$, followed by $H_M$, followed by $H_B$, followed by the weight-6 part of $H_V$.
    \item Next, starting from (odd,odd). Perform the weight-6 part of of $H_H$, followed by $H_M$, followed by $H_B$, followed by $H_V$ terms.
\end{itemize}
By this ordering in each of the last 4 steps we can optimize the unitary rotations that act on the same qubits and reduce the gates necessary in total from $20L$ $CX$ to $17L$ gates. The simplifications follow from the $CX$ gate cancellation when this ordering is used.

\bibliographystyle{apsrev4-1}
\bibliography{main}

\end{document}